\documentclass{IEEEcsmag}

\usepackage[colorlinks,urlcolor=blue,linkcolor=blue,citecolor=blue]{hyperref}

\usepackage{upmath}
\usepackage{amsmath}
\usepackage{background}
\usepackage{stackengine}

\jvol{XX}
\jnum{XX}
\paper{8}
\jmonth{May/June}
\jname{IEEE Intelligent Systems}
\pubyear{2019}

\SetBgContents{\normalsize\stackunder[36cm]{\Longstack{
    This~article~has~been~accepted~for~publication~in~IEEE~Intelligent~Systems.~This~is~the~author's~version~which~has~not~been~fully~edited~and\\ 
    content~may~change~prior~to~final~publication.~Citation~information:~DOI~10.1109/MIS.2021.3092768}}{%
    \rotatebox{0}{\Longstack{%
   ©~2021~IEEE.~Personal~use~is~permitted,~but~republication/redistribution~requires~IEEE~permission.~See~https://www.ieee.org/publications/rights/index.html~for~more~information.}}}\kern30pt}

\SetBgScale{0.7}
\SetBgAngle{0}
\SetBgOpacity{1}
\SetBgColor{blue}
\SetBgHshift{4cm}
\SetBgVshift{4.5cm}

\begin{document}

\sptitle{Department: Head}
\editor{Editor: Name, xxxx@email}
\title{Exploring Customer Price Preference and Product Profit Role in Recommender Systems}

\author{M. Kompan}
\affil{Kempelen Institute of Intelligent Technologies}

\author{P. Gaspar}
\affil{Slovak University of Technology in Bratislava}

\author{J. Macina}
\affil{Exponea}

\author{M. Cimerman}
\affil{Exponea}

\author{M. Bielikova}
\affil{Kempelen Institute of Intelligent Technologies}

\markboth{Department Head}{Paper title}

\begin{abstract}
Most of the research in the recommender systems domain is focused on the optimization of the metrics based on historical data such as Mean Average Precision (MAP) or Recall. However, there is a gap between the research and industry since the leading Key Performance Indicators (KPIs) for businesses are revenue and profit. In this paper, we explore the impact of manipulating the profit awareness of a recommender system. An average e-commerce business does not usually use a complicated recommender algorithm. We propose an adjustment of a predicted ranking for score-based recommender systems and explore the effect of the profit and customers' price preferences on two industry datasets from the fashion domain. In the experiments, we show the ability to improve both the precision and the generated recommendations' profit. Such an outcome represents a win-win situation when e-commerce increases the profit and customers get more valuable recommendations.
\end{abstract}

\maketitle

\chapterinitial{Recommender systems} are designed to help users with the information overload problem. One of the most common applications is e-commerce. The value of the recommender system for a business has been widely explored~\cite{Gomez-Uribe:2015:NRS:2869770.2843948,ieeeis}. Nowadays, recommender systems are an integral part of various well-known worldwide services (e.g.,~the Netflix claims that the combination of personalization and recommendation saves more than \$1B per year\footnote{\url{https://www.businessinsider.com/netflix-recommendation-engine-worth-1-billion-per-year-2016-6}}). Reflecting the success, more and more small to mid-size services are introducing recommenders (facing the problems of sparse data). Current approaches usually take into account past customer interaction to predict the future preferences of customers. The goal of the recommender system from the business perspective is not only to help customers to find relevant products or increase the loyalty of the customers but also to increase the revenue generated for the business. Therefore, a recommender system should optimize several objective functions at once.

A typical example of such an objective function is the profit of the recommended items or the profit generated by the recommender. The profit of an item is affected by the cost and the revenue~\cite{Walter2012}. While the cost is connected to the factors of purchasing a specific item by the business, revenue is understood as a means of distinguishing from other competitors (beyond the price) \cite{Walter2012}.

Since the profit of the items in the product catalog may vary, there might be several strategies of how it could be utilized in the recommendation, e.g., to recommend price-attractive products (even having low profit) to attract more customers or to recommend more profitable products to the customers that are not so sensitive to the prices.

One of the open research questions is how to incorporate a product profit into recommender systems
. In this paper, we aim at exploring the effect of manipulating the profit influence in the widely-used matrix factorization-based recommender (as a representative of collaborative filtering). 

Another essential perspective for the price of an item is the price preference of a customer. Analogously to the real world, various e-commerce brands focus on specific customer segments. However, we may find various price-sensitive customers within these segments as well. For this reason, we argue for the necessity to explore a model that considers customer price preference alongside product profit.

By manipulating the product profit and reflecting customer price preference, it can be cumbersome to retain or even increase the recommendation's precision. However, our results show that by combining the profit of recommended items and customer price preference, we can improve both the generated recommendations' profit and increase the recommendation precision. On the contrary, emphasizing the wrong profit and the customer's price preference level may lead to adverse effects, even customer churn. To explore this topic, we conducted an offline study on real-world datasets from the fashion domain (including real profit data). 

As a result, our approach can be easily incorporated into any score-based recommender system (e.g., matrix factorization). By manipulating our approach's hyperparameters, the e-commerce may influence (from the profit and price preference point of view) generated recommendations for each customer separately.
 
The main contributions of this paper are:
\begin{enumerate}
    \item Item profit and customer price preference easily applicable enhancement for Top-N recommender algorithms.
    \item The study of profit and price influence for the recommender system performance.
    \item Evaluation and study in real-world fashion e-commerce domains (including real-based profit data).
\end{enumerate}

\section{RELATED WORKS}
Several authors focused on the impact of recommender systems. Most of the studies indicate that incorporating a recommender to the e-commerce may result in a 30-35\% lift in the number of purchases\footnote{\url{https://www.mckinsey.com/industries/retail/our-insights/how-retailers-can-keep-up-with-consumers}}. 

Going further, profit optimization in the domain of recommender systems has been studied by several authors. Some of the works investigate how the profit can be incorporated into the recommendation and whether the profit impacts the quality of the recommendation. A related problem to profit optimization is the price level preference, which we usually need to consider, as well. As the main concerns about the potential customer loyalty losses are present, only a few studies try to explore the trade-off in real-field online settings~\cite{panniello}.

The topic of considering various objective functions is researched by the Multi-objective optimization
~\cite{Sener:2018:MLM:3326943.3326992}. Also, the constraint-based recommender systems may reflect various users preferences
~\cite{Felfernig2015}, e.g., price level preference.

Anderson studied a relationship between customer satisfaction and price tolerance~\cite{Anderson1996}. It was found that the increase in customer satisfaction can be associated with a decrease in price sensitivity. In other words, price tolerance is an indicator of the satisfaction of the customers.

The price level preference of e-commerce customers varies across the product categories. Clearly, such preference may be modeled and, in the next step, predicted to improve the recommendations~\cite{azaria2013movie}. The study \cite{jannach2017determining} showed that recommending products from the same price level within a session resulted in a doubled increase of recommendation-attributed conversion rate.

Chen et al. transformed product prices into a particular customer price utility. Next, they described customer price preferences within a category by a distribution over price utility~\cite{Chen:2016:BRU:3001595.2978579}. Finally, they fused two matrix representations of customer price preferences to improve recommendation accuracy. 


Another direction towards the profit awareness recommender was explored in~\cite{chen2008developing}. The two most notable recommenders were Convenience Plus Profitability Perspective Recommender System (CPPRS) and Hybrid Perspective Recommender System (HPRS). CPPRS recommends products based on both purchase probability and product profitability, whereas HPRS applies similarity-based purchase profitability and a probability approach. Experimental results showed that HPRS could improve profitability while preserving model accuracy. This method is rather suitable for short-term profitability gains in cross-selling use-cases. 

Louca et al. explored objective function that optimized the likelihood of the purchase, and the profit of item~\cite{Louca}. The preliminary results on the real-world dataset indicated that the profit of generated recommendations increased; however, the precision of such recommendations was lower than the baseline (logistic regression).

Azaria et al. explored the profit on a utility maximizer algorithm that operated on a ranked list generated by a black-box recommender system~\cite{azaria2013movie}. Their algorithm learned a model as a function of both the movie rank and the price, which resulted in a likelihood of customers willing to watch a movie. The authors showed a clear conflict between the business and customer goals, which leads to a reduced purchase probability of a customer. Wang et al. contributed~\cite{wang2009mathematical} by developing a comprehensive mathematical, analytical model. The numerical example demonstrated three possible market strategies: Profit Strategy, Win-Win, and Global Optimal Strategy. Even though one can focus on a Win-Win strategy and therefore minimize the loss of customers, in the long run, there is no explicit strategy for a long-term benefit for both businesses and customers.




The long-term profit optimization was also explored by Hosein et al.~\cite{Hosein}. Authors suggest considering both short- and long-term profit. The proposed approach uses several assumptions, which may restrictive in the real-world application - e.g. - a suitable product is always available for a customer. The evaluation of the proposed approach was performed over the simulated profit, which, in fact, is far away from the real-world application.

Pei et al. proposed a value-aware recommender system that maximized the expected profit of a recommendation list~\cite{Pei:2019:VRB:3308558.3313404}. To achieve this, they assigned an economic value to several types of customer actions (e.g.,~add to a basket, add to a wishlist). The authors considered a profit to be a sum of sales in a specific time frame. They utilized profit optimization using an evolution strategy, where the reward was an expected profit. The proposed model learned the customer state after each action, where several features were considered (such as customer age, gender, purchase power, item ctr, price). Based on the offline and online experiments, they found that their method improved the precision, recall, MAP, and expected profit of the recommendation lists.

The alternative to incorporating profit into recommenders is to generate personalized prices of products. In fact, similar aspects have to be considered. Kamishima and Akaho proposed a multi-armed bandit algorithm that considers three types of customers: standard, discount oriented, and indifferent~\cite{Kamishima:2011:PPR:2039320.2039329}. Based on these types, a reward is distributed. However, the usage of personalized prices is still a controversial approach that rises many ethical questions and is usually not explored by academia. 
 
To sum it up, there is an increasing trend in incorporating a business perspective into the recommender systems, which results in a higher activity of academia. The research is usually focused on multi-objective optimization, which requires a new recommender to be designed. These approaches are often unsuitable for a general application to average e-commerce, as they require high computation power or large volumes of data (which are not available in small to mid-size e-commerce).

The multi-objective optimization often fails to clearly evaluate particular objectives - price preference level, profit, and precision. In this paper, we aim to explore these effects and bring some insights into the relationship of these metrics (i.e., objectives). Based on our conclusions, more transparent and trustful methods can be proposed in the future. 

Moreover, published studies often fail at the evaluation stage, as the profit is unavailable for most of the public datasets. As a result, there is no clear indication of whether both the precision and the profit of generated recommendations can be increased. We can expect that the profit is explored in more detail by the business. Such results are not usually presented to the public. They often may be used for improving the short-term revenues without considering the customer's utility functions.


\section{PRICE PREFERENCE AND PROFIT ENHANCED RECOMMENDATIONS}
In order to be able to extend any scoring-based baseline recommender, we propose a new scoring function. We have several constraints that need to be reflected. The proposal follows the idea of designing a price preference- and profit-aware extension of a typical recommender used in the e-commerce domain (which uses a score to sort the candidates). Moreover, we aim to propose a method that would allow us to control the impact of its two main components -- the profit and the price preference of a customer. As a result, the business can control the recommender's outcome by simply adjusting the parameters, which will result in recommending products with low profit and vice versa. Besides, based on lessons learned in real e-commerce analysis, several constraints should be reflected, e.g., to not completely disrupt the original recommender score or products with negative profit.

Let $\widehat{r}_{ui}$ be a predicted score (generated by arbitrary baseline recommender, normalised to $\left \langle 0,1 \right \rangle$ interval) for the customer $u$ and the item $i$. Based on the described idea, we propose the profit and price preference-aware score as:

\begin{equation}
\small
\begin{split}
    s_{ui}=\left [ 1 + \log_{10} \left ( 0.1+\frac{0.9 * retail\: price_i}{price_i} \right ) \right ]^{\alpha}+\\
   + \left [ 1+\log_{10} \left ( 0.1+\frac{0.9 * retail\: price_i}{\overline{retail\: price_u}} \right ) \right ]^{\beta}
\end{split}
\end{equation}

\noindent where $retail\; price_{i}$ is a selling price of an item $i$ for the customers, and $price_i$ is the purchase price of an item $i$ (purchased by the retailer). The $\overline{retail\: price_u}$ is the average selling price of products bought by the customer $u$. The recommended item's profit is captured as the ratio of $\frac{retail\: price_i}{price_i}$. Similarly, the difference between a customer's typical price level and the actual item's price is captured as $\frac{retail\: price_i}{\overline{retail\: price_u}}$. The two hyperparameters allow us to control and to manipulate the effect of the profit and price preference influence:
\begin{itemize}
    \item $\alpha$ - adjusts the influence of the profit ($\alpha=0$ no influence, $\alpha=1$ prefer items with higher profit, $\alpha=-1$ prefer item s with a lower profit).
    \item $\beta$ - controls the influence of the price preference part ($\beta=0$ no influence,  $\beta=1$ prefer more expensive, $\beta=-1$ cheaper items (in comparison to average price of items bought by the customer)).

\end{itemize}
We use the $\log_{10}$ decay factor for both parts - the profit and price preference. Thanks to this factor, we do not disrupt the original score and reflect the variability of prices usually used in retail. Similarly, we adjusted the logarithm function to return only positive values and to be defined for products with any profit (including a negative profit, i.e., products used in marketing campaigns). We expect both the retail price and price to be non-zero positive values.

The adjustment of the baseline predictions can be easily done by just multiplying the original ($\widehat{r}_{ui}$) and profit-aware scores ($s_{ui}$) (Equation~2). The Top-N recommender then recommends items with the highest $\widehat{p}_{ui}$. 
\begin{equation}
   \widehat{p}_{ui}=s_{ui} * \widehat{r}_{ui}
\end{equation}
As the proposed enhancement is used to adjust the original recommender score, it can be easily implemented to existing recommendation processes. Moreover, setting the hyperparameters $\alpha,\beta$ to $0$ results in obtaining the original score of the arbitrary recommender (i.e.,~assigns the weight of the profit and price preference awareness to zero).

To sum it up, in this manner, we can apply the proposed profit and price preference-aware adjustment (Equation~1) to any existing recommender, which produces a score prediction, by a simple multiplication step (Equation~2). Moreover, by manipulating the hyperparameters (set manually or automatically by hyperparameter search), items with both lower or higher profits can be recommended. In this way, the recommender may be used as an active marketing tool.


\section{EVALUATION}
We performed an offline study to explore the effect of the profit and the price preference based on the proposed scoring function in the generated recommendation. We investigated the impact of the product's profit on the recommendation performance considering traditional metrics (i.e.,~Precision, Recall, MAP) and a business metric (i.e.,~Profit at Hit). The evaluation aimed to explore various settings (i.e.,~hyperparameters) and their effects on the item's rank (including precision).

\subsection{Dataset}
The study was performed over two e-commerce sites in a fashion domain\footnote{Data were obtained from the Exponea customer data platform: https://exponea.com}(exclusively fashion-based product catalog). We used a representative sample of the customers' activity. This results in approx. 500k customers and 4.8M actions for the Dataset~1; and 180k customers and 1.2M actions for the Dataset~2. There were two types of customer actions available: displaying the product detail and purchase of the product. We considered the purchase action as an indicator of the positive implicit feedback (true positive). The datasets differ in the average number of interactions and purchases (Table~\ref{tab01}). Moreover, Dataset~2 contains products with obviously lower profit and lower customers interactions. Dataset~2 is an example of an e-commerce site targeting low-income customers. 

\begin{table}[]
\caption{Descriptive statistics of datasets (DS1 and DS2) that were used for the study from the fashion domain.}
\centering
\begin{tabular}{|l|c|c|}
\hline
\textbf{Characteristic}                      & \textbf{DS 1} & \textbf{DS 2} \\ \hline
Avg. no. of interactions per customer   & 9.46      & 6.45      \\ \hline
Avg. no. of purchases per customer      & 6.00      & 2.78      \\ \hline
Customers with less than 3 actions & 48.6\%    & 54.5\%    \\ \hline
Avg. profit of products             & 161\%     & 87\%     \\ \hline
Median profit of products           & 169\%     & 75\%     \\  \hline
No. of unique customers           & 500k     & 180k     \\  \hline
No. of actions           & 4.8M     & 1.2M     \\  \hline
No. of unique products           & 50k     & 49k     \\  \hline
75th percentile of actions           & 8     & 6     \\  \hline
95th percentile of actions           & 39     & 24     \\  \hline
\end{tabular}
\label{tab01}
\end{table}

\subsection{Baseline}
The idea of the proposed study bases on a baseline approach that generates a score (e.g., rating, probability of a purchase). This score is then adjusted to reflect the profit of the recommended items and the customer's price preference. Based on this constraint, we utilize a standard, widely-used, matrix factorization model~\cite{DBLP:conf/recsys/Kula15}.
Such an arbitrary score-based recommender allows us to demonstrate the broad application of the proposed extension, respectively, allowing us to explore both elements' effects. The prediction for a customer $u$ and item $i$ is calculated as follows:

\begin{equation}
  \widehat{r}_{ui} = f(\mathbf{q_u}\cdot\mathbf{p_i}+b_u+b_i)
\end{equation}
where $q_u$ is customer latent representation and $b_u$ customer bias; $p_i$ is item latent representation and $b_i$ is item bias; and $f(\cdot)$ is a sigmoid function\footnote{The full details are available in the Kula's original paper~\cite{DBLP:conf/recsys/Kula15}.}. For the training procedure, a WARP loss function was used~\cite{37180}, with the following training parameters (based on the parameter optimization): learning rate - 0.05, number of latent components - 50 and the number of optimization epochs - 50.

\subsection{Methodology}
To explore the effect of both hyperparameters proposed in Equation~1, we generated recommendations based on the baseline approach. We split each dataset (users' activity) into the train and the test set based on the time with 80:20 ratio. 

For each dataset, we generated recommendations using the baseline model. For all experiments the purchase action is considered as ''hit" or ''true positive". Predicted ratings were adjusted based on the Equation~1 and Top-10 items with the largest $\widehat{p}_{ui}$ were recommended to each customer (Equation~2). We computed the $\widehat{p}_{ui}$ score for all the hyperparameters combinations, with the step $0.1$ and the ranges: $\alpha \in \left \langle -1,1 \right \rangle$ and $\beta \in \left \langle -1,1 \right \rangle$. 

Next, we computed the Precision@k, Recall@k, and MAP@k metrics 
with $k=10$. 


On the contrary, we observed the profit of the generated recommendations by the \textit{Profit at Hit} metric. We introduced the metric to capture the profit of the items that the customer bought. The PAH@k is computed as follows:
\begin{equation}
    PAH@k=\frac{\frac{\sum_{i=1}^{k} profit(True Positive)}{\left | True Positive \right |}}{\left | U \right |}
\end{equation}

\noindent where $TruePositive$ are the correctly recommended (and purchased) items, and $U$ represents the set of the users in the test set. In this manner, we guarantee to consider only items that were interesting for the customer.

\subsection{Results}
The intuition behind the profit-aware recommender system is that the items with low profit would be more attractive for the customers (as the item's real value is closer to the market price). Similarly, negative profit can be used as a marketing tool. On the contrary, the market theory supports the opposite \cite{econ}, i.e., more popular and thus demanded items will result in higher profit levels. 


\begin{figure*}[]
\centering
\includegraphics[width=0.985\textwidth]{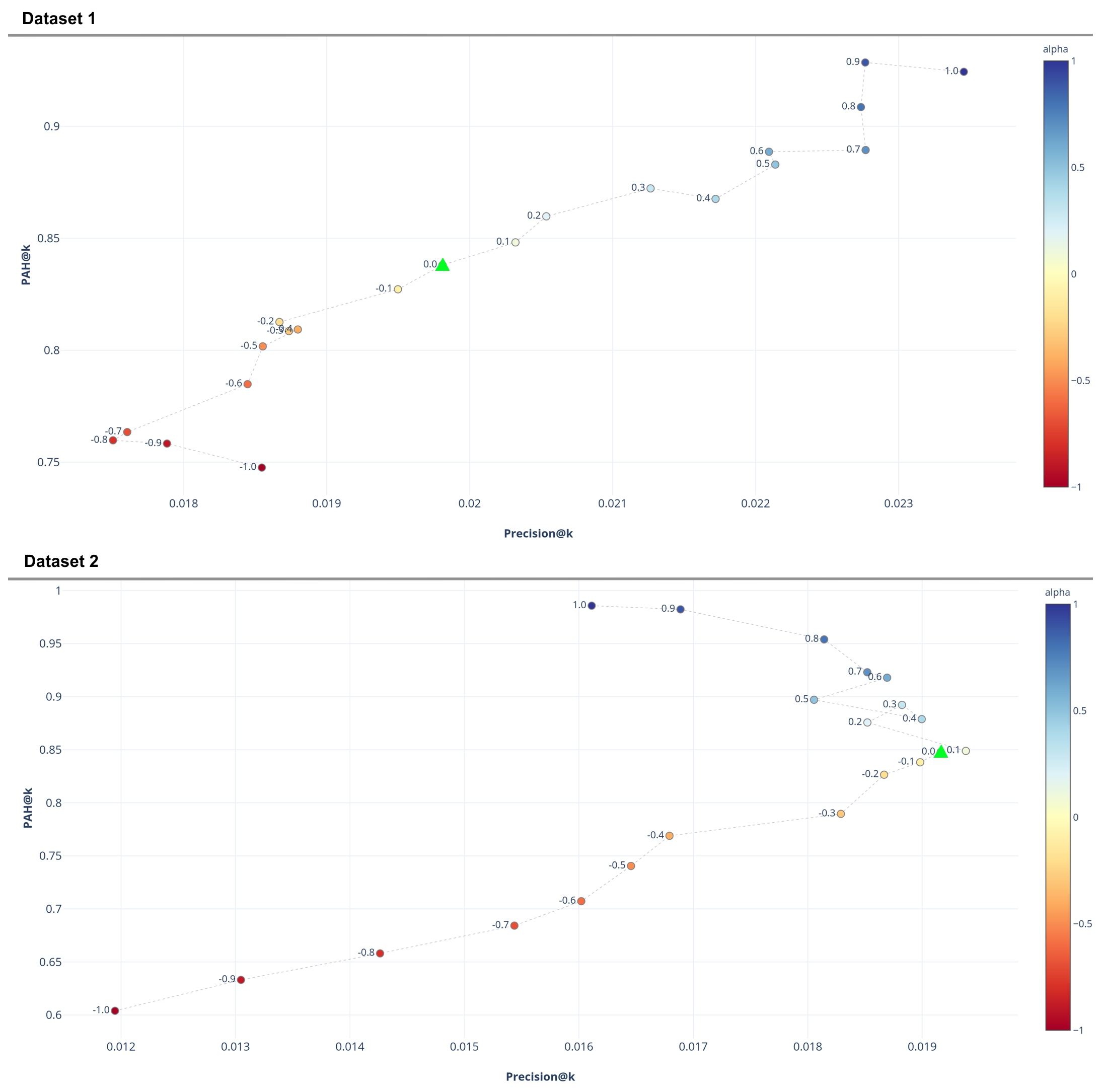}
\caption{The effect of the $\alpha$ parameter (the impact of the profit-awareness) on the Precision and PAH metrics. The positive values of $\alpha$ (blue color) prioritize items with higher profit and vice versa (negative values of $\alpha$ red color). The Dataset~1 - top, Dataset~2 - bottom. The green triangle represents the baseline recommender which does not consider the profit-awareness.}
\label{fig:alpha}
\end{figure*}

\subsubsection{Profit-awareness}
We believe that the profit-aware recommender has to consider the trade-off between the profit and the recommendation precision. As we can see (Figure~\ref{fig:alpha}), the hyperparameter $\alpha$ has a significant impact on both the precision and the profit. The negative values ($\alpha \in \left \langle -1,0 \right \rangle$) hurt the profit metrics as expected (negative $\alpha$ prefers items with lower profit). Surprisingly, the precision metric is also lower than the baseline. This indicates that customers on average, are not willing to buy products of certain characteristics (as we believe these characteristics are reflected in the price and the profit, respectively). From another perspective, this may be caused by the unfamiliarity with some product range. However, e-commerce sites used in our datasets do not consider the price or profit of items, and thus no explicit bias is present in the data. The Dataset~1 and Dataset~2 have slightly different results for the positive values of $\alpha$ (Figure~\ref{fig:alpha}). We believe that this is caused by the different characteristics of the e-commerce site (i.e., dataset), which in fact aims at discounted products (Dataset~2). The highly profitable products in such e-commerce are generally in lower demand from its nature.

The Profit at Hit metrics (PAH) is logically increased as more profitable products are recommended (positive values of $\alpha$). We want to point out that we were able to increase the precision and the profit for both datasets, respectively (Figure~\ref{fig:alpha} - points right and above in respect to the triangle). This is extremely important to e-commerce, as no trade-off needs to be considered, and an optimal solution for e-commerce and a customer is found. We were able to improve the profit of generated recommendations simultaneously with precision.

\begin{figure*}[]
\centering
\includegraphics[width=0.988\textwidth]{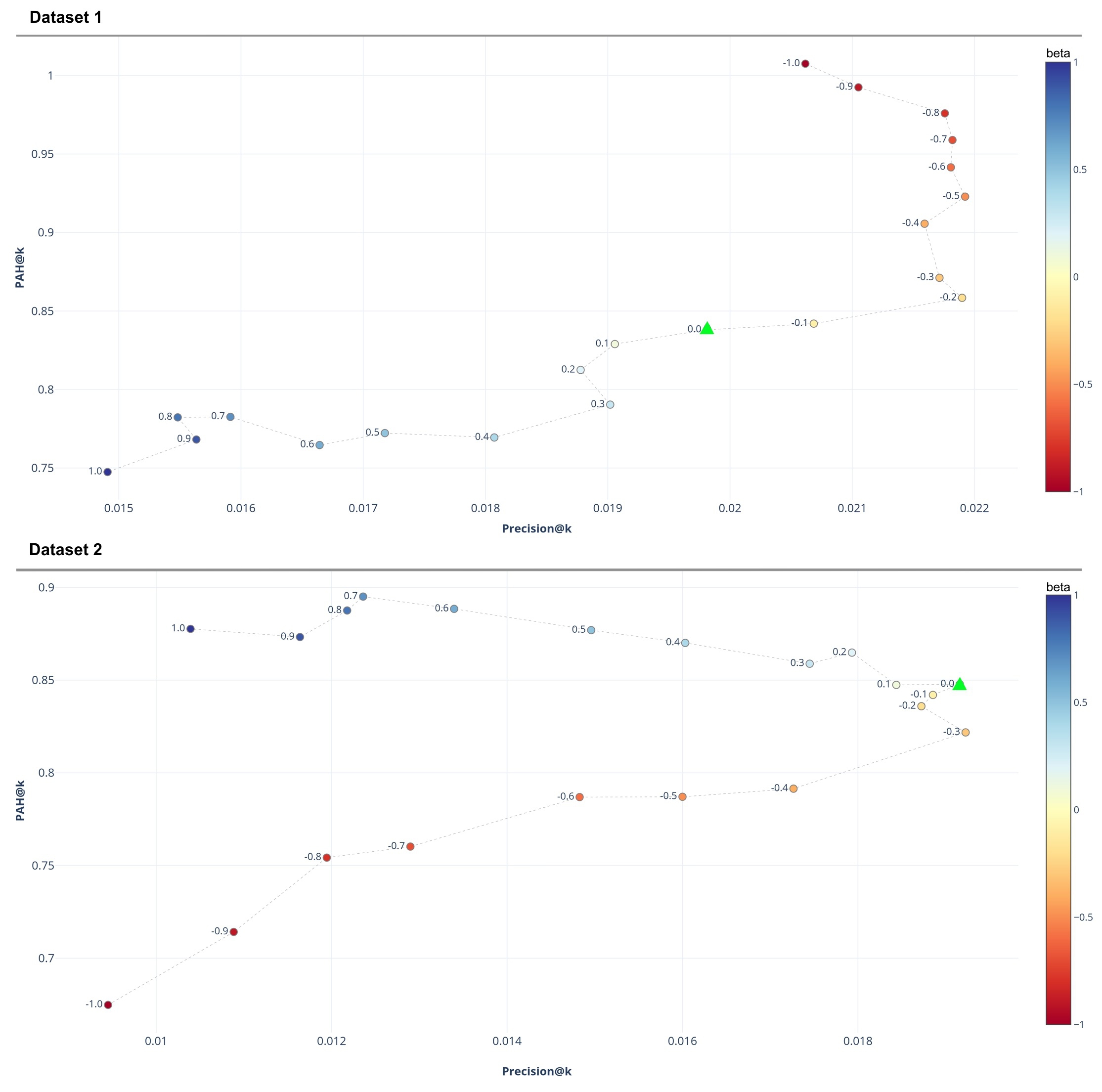}
\caption{The effect of the $\beta$ parameter (the impact of the price preference-awareness) on the Precision and PAH metrics. The positive values of $\beta$ (blue color) prioritize more expensive items (as customer's average) and vice versa (negative values of $\beta$ - red color). The Dataset~1 - top, Dataset~2 - bottom. The green triangle represents the baseline recommender which does not consider price preference-awareness.}
\label{fig:beta}
\end{figure*}

Another conclusion based on the results (Figure~\ref{fig:alpha}) indicates that e-commerce can adjust the recommender on a specific use-case (i.e., marketing campaign). By setting the $alpha$ parameter to prefer, i.e., recommend more or less profitable products. Moreover, various $\alpha$ values can be set for various customers.

\subsubsection{Price preference-awareness}
Next, we explored the impact of considering price preference awareness. As expected, positive $\beta$ values (Figure~\ref{fig:beta}) result in higher Profit at Hit for the Dataset~2 (the profit is computed strictly on purchased items only). Hand by hand with the PAH increases the Precision decreases. This is a straightforward result. In other words, if we recommend more expensive items, the profit increases, and the precision (i.e., the number of purchases) decreases. An inverted pattern can be observed when the $\beta$ values are negative, i.e., we recommend cheaper items. The PAH logically decreases; however, the precision decreased either. This pattern can be explained, in our opinion, by highly profiled users, which tend to buy similar brands (at a specific price level).

\begin{figure*}[ht!]
\centering
\includegraphics[width=0.99\textwidth]{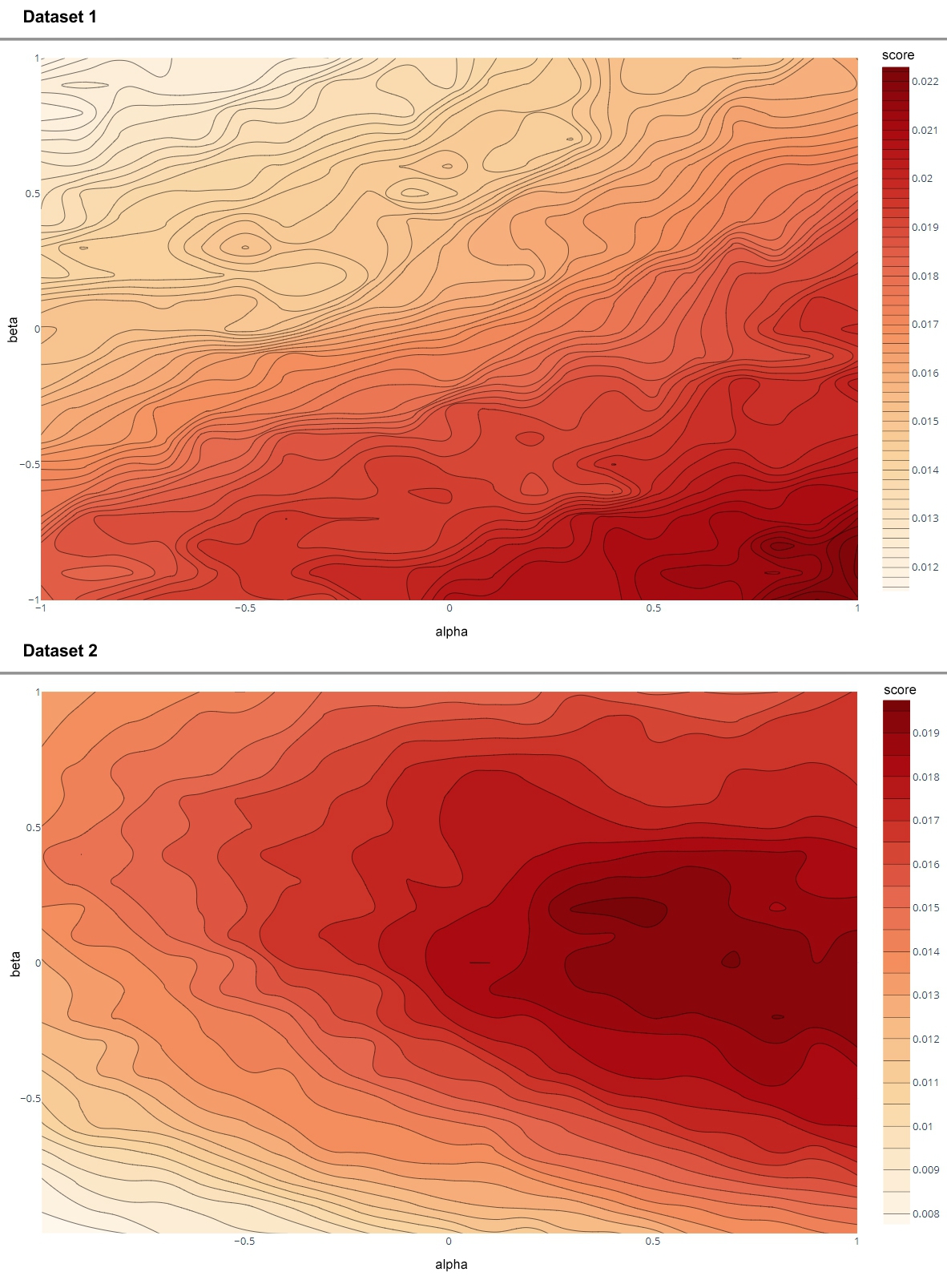}
\caption{The proposed profit and price preference-aware score for the Dataset~1 - top, Dataset~2 - bottom. Each combination of $\alpha$ and $\beta$ is presented. The color intensity combines the profit and the precision of recommendations (darker is better).}
\label{fig:merged}
\end{figure*} 

The most interesting result occurs for Dataset~1 (Figure~\ref{fig:beta}). By recommending cheaper products (i.e., negative $\beta$) both PAH and the Precision increased. This can be explained by the different customer segments present. The collaborative filtering used as a baseline tends to recommend a mix of various priced products. When considering each customer's price preference separately, we can increase the number of correctly recommender items. The increase suppresses the lower profit of such items and results in higher PAH. This is supported by the Precision metric increase.

\subsubsection{Combined score}
Finally, we combined both parts of the score (the profit and price preference awareness). As we can see (Figure~\ref{fig:merged}) for both datasets, the proposed enhancement was able to generate more precise and also more profitable recommendations (Table~\ref{tab_res}) at the same time. This outperformed our expectations, as we aimed at exploring the effect of both hyperparameters. In general, the proposed enhancement allows us to adjust the recommendations according to a retailer's needs. Since our experiments are performed offline, even more impact on both metrics is expected in a real application -- where the generated recommendations may influence the customers.

\begin{table}[]
\caption{Results obtained by the baseline matrix factorization model and the profit-aware recommender with the best hyperparameter settings. The best settings are selected based on the Precision@10 metric (P@10). For the comparison also the Recall@10 (R@10), Mean Average Precision (MAP@10), and the Profit at Hit (PAH@10) metrics are presented.}
\centering
\begin{tabular}{lcccc}
\hline
\multicolumn{5}{|c|}{\textbf{Dataset 1}} \\ \hline
\multicolumn{1}{|l|}{}             & \multicolumn{1}{l|}{P@10} & \multicolumn{1}{l|}{R@10} & \multicolumn{1}{l|}{MAP@10} & \multicolumn{1}{l|}{PAH@10} \\ \hline
\multicolumn{1}{|l|}{Baseline}     & \multicolumn{1}{c|}{.0198}   & \multicolumn{1}{c|}{.0197}  & \multicolumn{1}{c|}{.0157}      & \multicolumn{1}{c|}{.8380}                      \\ \hline
\multicolumn{1}{|l|}{Profit aware} & \multicolumn{1}{c|}{.0245}   & \multicolumn{1}{c|}{.0244}  & \multicolumn{1}{c|}{.0193}      & \multicolumn{1}{c|}{1.065}                      \\ \hline
\multicolumn{5}{|c|}{\textbf{Dataset 2}}                                                                                                                             \\ \hline
\multicolumn{1}{|l|}{Baseline}     & \multicolumn{1}{c|}{.0191}       & \multicolumn{1}{c|}{.0190}      & \multicolumn{1}{c|}{.0115}      & \multicolumn{1}{c|}{.8477}                          \\ \hline
\multicolumn{1}{|l|}{Profit aware} & \multicolumn{1}{c|}{.0196}       & \multicolumn{1}{c|}{.0195}      & \multicolumn{1}{c|}{.0120}      & \multicolumn{1}{c|}{.8774}                          \\ \hline
\end{tabular}
\label{tab_res}
\end{table}


\section{CONCLUSIONS}
The profit of the generated recommendations is one of the essential utility functions from the business perspective. However, this is rarely reflected by the researchers. To explore this topic, we proposed a profit and price preference-aware enhancement for the standard score-based recommenders (by adjusting baseline scores). In this manner, the proposed approach applies to most recommender systems used in the small and medium-size e-commerce business. 

Our proposal aims to include the profit of recommended items and the customer's price preference in the recommendation process. The hyperparameters can control each part's effect, allowing us to adjust generated recommendations concerning the business needs and/or specific customer groups.  

We explored the effect of both hyperparameters in the study on two real e-commerce datasets (from a fashion domain), which is rare in the profit-aware recommender system domain. By manipulating the proposed enhancement's hyperparameters (Equation~1), we can adjust the recommendations to both lower or higher profits. As we showed, even a standard widely-used recommender can be easily extended to consider the profit of the generated recommendations and the customer's price preference. Furthermore, we showed that the proposed extension could adapt to datasets with various characteristics. Moreover, we were able to increase the profit of the true positives and the precision of such recommendations.

On the contrary, the proposed approach allows us to adjust generated recommendations in both directions (i.e., to recommend items with lower profit), which can be further used to target specific customer groups. This is a promising result, which indicates further research that can result in benefits for customers and e-commerce. Going further, the comparison of results with online evaluation outcomes may implicate exciting findings in a broader evaluation philosophy context. Last but not least, the outcomes of our and similar studies may help to further propose more transparent and trustful methods in the recommender systems fields.

\section{ACKNOWLEDGMENT}
This research was partially supported by the COST Action 19130. It is a result of the joint research lab by Exponea.

\begin{IEEEbiography}{M. Kompan}{\,}is an associate professor and senior researcher with the Kempelen Institute of Intelligent Technologies, Bratislava, Slovakia. His research interest are among recommender systems, user modelling and prediction methods. He is a member of the IEEE.
\end{IEEEbiography}

\begin{IEEEbiography}{P. Gaspar}{\,}is a PhD. student with the Slovak University of Technology in Bratislava. In his research, he explores the topics of context-aware recommender systems.
\end{IEEEbiography}

\begin{IEEEbiography}{J. Macina}{\,}is a lead machine learning engineer with the Exponea. His research background is focused in the field of machine learning, information retrieval, and recommender systems. 
\end{IEEEbiography}

\begin{IEEEbiography}{M. Cimerman}{\,}is a machine learning engineer with the Exponea. In his work, he is addressing the challenges and problems in recommender systems, machine learning and data science.
\end{IEEEbiography}

\begin{IEEEbiography}{M. Bielikova}{\,}is a professor and senior researcher with the Kempelen Institute of Intelligent Technologies, Bratislava, Slovakia. She is focusing on artificial intelligence, particularly in human computer interaction modelling, machine learning, user modeling. She is active in discussions related to ethics in IT and AI. She is a senior member of the IEEE.
\end{IEEEbiography}

\end{document}